\documentclass[a4paper,11pt]{article}
\usepackage{pos}

\usepackage[utf8]{inputenc}
\usepackage[T1]{fontenc}
\usepackage[english]{babel}
\usepackage{graphicx}
\usepackage{xcolor}
\usepackage{mathtools}
\usepackage{siunitx}
\usepackage[final]{microtype}
\usepackage{hyperref}

\hypersetup{%
    pdftitle = {Critical Endpoint of QCD and Baryon Number Fluctuations in 
    a Finite Volume},
    pdfauthor = {Julian Bernhardt, Christian S. Fischer, Philipp Isserstedt},
    bookmarksopen = true,
    colorlinks = true,
}

\graphicspath{{./Graphics/}}

\newcommand*{\+}{\hspace*{.08335em}}

\newcommand{\ii}{\ensuremath{\mathrm{i}}}
\newcommand{\dd}{\ensuremath{\mathrm{d}}}

\newcommand{\upu}{\ensuremath{\mathrm{u}}}
\newcommand{\upd}{\ensuremath{\mathrm{d}}}
\newcommand{\ups}{\ensuremath{\mathrm{s}}}
\newcommand{\upB}{\ensuremath{\mathrm{B}}}
\newcommand{\upQ}{\ensuremath{\mathrm{Q}}}
\newcommand{\upS}{\ensuremath{\mathrm{S}}}

\newcommand{\muu}{\ensuremath{\mu_{\upu}}}
\newcommand{\mud}{\ensuremath{\mu_{\upd}}}
\newcommand{\mus}{\ensuremath{\mu_{\ups}}}

\newcommand{\bbZ}{\ensuremath{\mathbb{Z}}}

\DeclareMathOperator{\Tr}{Tr}
\DeclarePairedDelimiterX{\expval}[1]{\langle}{\rangle}{#1}
\renewcommand{\bar}{\overline}

\newcommand{\UA}{\textup{U}_{\textup{A}}}

\DeclareSIUnit{\MeV}{\mega\electronvolt}
\DeclareSIUnit{\GeV}{\giga\electronvolt}
\DeclareSIUnit{\fm}{\femto\metre}

\title{%
    Critical Endpoint of QCD and Baryon Number Fluctuations in a Finite Volume
}
\ShortTitle{%
    Critical Endpoint of QCD in a Finite Volume
}

\author*[a,b]{Julian Bernhardt}
\author[a,b]{Christian S. Fischer}
\author[a,b,\dagger]{Philipp Isserstedt}

\affiliation[a]{%
    Institut f\"{u}r Theoretische Physik, %
    Justus-Liebig-Universit\"{a}t Gie\ss{}en, %
    35392 Gie\ss{}en, %
    Germany%
}

\affiliation[b]{%
    Helmholtz Forschungsakademie Hessen f\"{u}r FAIR (HFHF), %
    GSI Helmholtzzentrum f\"{u}r Schwerionenforschung, %
    Campus Gie\ss{}en, %
    35392 Gie\ss{}en, %
    Germany%
}

\notes{%
    \note[$\dagger$]{%
        Present address:
        Frankfurt Consulting Engineers,
        Bessie-Coleman-Str.~7,
        60549 Frankfurt am Main,
        Germany.
    }
}

\emailAdd{julian.bernhardt@physik.uni-giessen.de}
\emailAdd{christian.fischer@theo.physik.uni-giessen.de}
\emailAdd{philipp.isserstedt@physik.uni-giessen.de}

\abstract{%
    We summarize recent results on the volume dependence of the location of the
    critical endpoint in the QCD phase diagram. To this end, we employ a
    sophisticated combination of Lattice Yang--Mills theory and a (truncated)
    version of Dyson--Schwinger equations in Landau gauge for $2 + 1$ quark
    flavours. We study this system at small and intermediate volumes and
    determine the dependence of the location of the critical endpoint on the
    boundary conditions and the volume of a three-dimensional cube with edge
    length $L$. We also discuss recent results on baryon number fluctuations in 
    this setup.
}

\FullConference{%
    FAIR next generation scientists - 7th Edition Workshop (FAIRness2022)\\
    23-27 May 2022\\
    Paralia (Pieria, Greece)
}

\begin{document}

\maketitle

\section{\label{sec:intro}%
    Introduction
}
\vspace{-1mm}

One of the prime goals of many heavy-ion collision experiments is to verify the existence and
determine the location of the putative critical endpoint (CEP) in the phase
diagram of QCD. However, the hot and dense QCD matter formed in these
experiments is only finite in spatial extent, which depends, e.g., on the size
of the colliding ions and the centrality of the collision. As a consequence, it
serves as an important cross-check between theory and experiment to analyse
potential dependences of the experimental signatures on this finite volume in
theoretical calculations. To this end, it is especially interesting to
investigate the fluctuations of conserved charges such as the baryon number.

In the following, we summarize our findings for the QCD phase diagram in a
finite volume obtained from Dyson--Schwinger equations (DSEs) as found in
Ref.~\cite{Bernhardt:2021iql}. Additionally, we present very recent results from
Ref.~\cite{Bernhardt:2022mnx} for the finite-volume baryon number fluctuations
around the CEP in the same framework.

\section{\label{sec:dses}%
    Dyson--Schwinger Equations
}

For all following investigations, the main quantity of interest is the
nonperturbative quark propagator $S_{f}$ of quark flavour $f \in\{\upu/\upd,
\ups\}$ -- i.e., a $2+1$ flavour setup -- at temperature $T$ and quark chemical 
potential $\mu_{f}$. It is
calculated using DSEs which are the quantum equations of motion for the
correlation functions of a quantum field theory. While in principle not
requiring any approximations, they imply an infinite tower of coupled equations
that has to be truncated for practical calculations.

We use a truncation that has been used extensively for investigations of the QCD
phase diagram in the past (see, e.g., our recent works in
Refs.~\cite{Isserstedt:2019pgx, Gunkel:2019xnh, Isserstedt:2020qll,
Gunkel:2021oya, Bernhardt:2021iql, Bernhardt:2022mnx}). That is, we solve the
DSEs for the quark and gluon propagators using a model ansatz for the
quark--gluon vertex. For the gluon, we employ quenched lattice data and take
unquenching effects into account by calculating the quark loop explicitly. For
the sake of brevity, we refer the reader to the review~\cite{Fischer:2018sdj}
and references therein for explicit expressions and more details.

\section{\label{sec:volume}%
    Finite-Volume Framework
}

In order to specify a three-dimensional finite-volume framework, we need to fix
both a shape and the boundary conditions of the volume. A mathematically
feasible choice is a cube with edge length $L$ and antiperiodic/periodic spatial
boundary conditions (ABC/PBC). As a consequence, each component in momentum
space can only assume discrete values of the spatial Matsubara modes,
\begin{equation}
    \omega_{n}^{L}
    =
    \frac{\uppi}{L}
    \times
    \begin{cases}
        2n
        &
        \text{for PBC}
        \,,
        \\
        2n + 1
        &
        \text{for ABC}
        \,,
    \end{cases}
    \quad n \in \bbZ
    \,,
\end{equation}
and all momentum integrals in our DSEs become sums, which
is the central relation of our setup:
\begin{equation}
    \int \frac{\dd^{3} q}{(2 \uppi)^{3}}
    \,
    K(\vec{q})
    \, \to \,
    \frac{1}{L^{3}} \sum_{\vec{n} \in \bbZ^{3}}
    K(\vec{q}_{\vec{n}})
    \,,
    \
    \text{ with }
    \
    \vec{q}_{\vec{n}}
    :=
    \sum_{i = 1}^{3}
    \omega^{L}_{n_{i}}\vec{e}_{i}
    \,,
\end{equation}
for a generic integrand $K$. For practical calculations, however, more
intricacies have to be taken care of which is described in detail in
Ref.~\cite{Bernhardt:2021iql}.

\section{\label{sec:results}%
    Results
}

\paragraph{QCD Phase Diagram}

We start our investigation of the QCD phase diagram with the volume dependence
of crossover line and CEP as performed in Ref.~\cite{Bernhardt:2021iql}. To this
end, we employ the quark condensate as the order parameter for chiral symmetry
breaking which in our setup reads
\begin{equation}
    \expval{
        \bar{\psi} \psi
    }_{f}
    \sim
    -
    \frac{T}{L^{3}}
    \sum_{\omega_{m}}
    \sum_{\vec{n} \in \bbZ^{3}}
    \,
    \Tr\bigl[
        S_f(\omega_{m} + \ii \mu_{f}, \vec{q}_{\vec{n}})
    \bigr]
    \,,
\end{equation}
where $\omega_{m}$ are the temporal Matsubara frequencies introduced by the
temperature. For any nonvanishing bare quark mass, the condensate is divergent
and needs to be regularized. To this end, we define the subtracted condensate,
$\Delta_{\upu\ups} := \expval{\bar{\psi} \psi}_{\upu} - (m_{\upu} / m_{\ups})
\expval{\bar{\psi} \psi}_{\ups}$. The pseudocritical temperature of the chiral
crossover transition is in our case defined as the inflection point of
$\Delta_{\upu\ups}$ with respect to temperature, while the CEP is the point
where the crossover becomes second order.

\begin{figure*}[t]
    \centering%
    \includegraphics[scale=1.0]{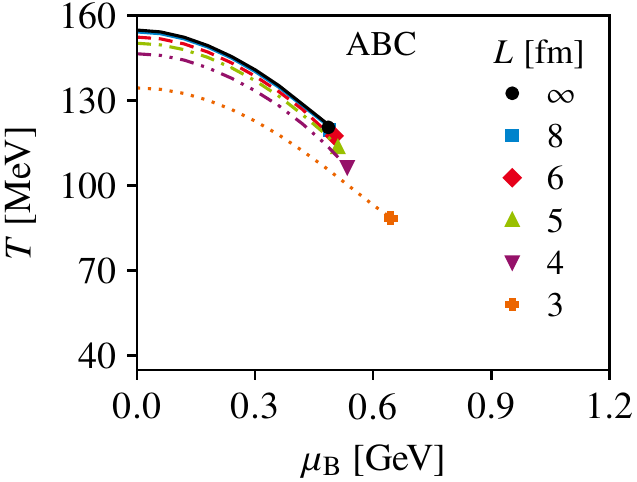}%
    \hspace*{1em}
    \includegraphics[scale=1.0]{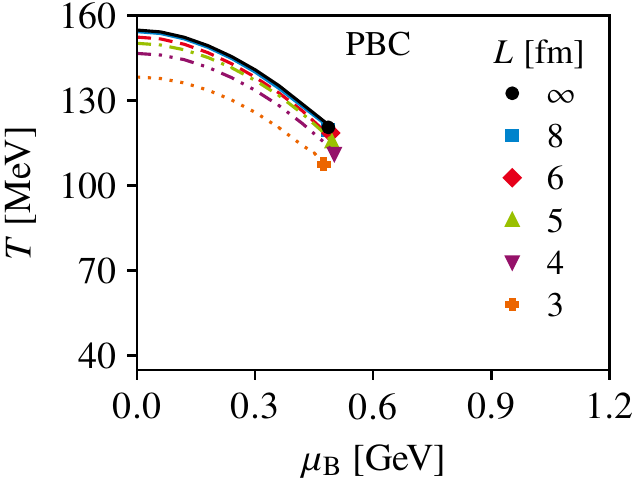}%
    \vspace*{-1mm}%
    \caption{\label{fig:CEP}%
        Finite-volume effects on the QCD phase diagram for antiperiodic
        (\emph{left}) and periodic (\emph{right}) boundary conditions and
        different box sizes. Here, we show the crossover lines and locations of
        the CEPs (symbols).
    }%
\end{figure*}

In Figure~\ref{fig:CEP}, we show our findings for the QCD phase diagram for both
boundary conditions and systems sizes between $L = \SI{3}{\fm}$ and $L =
\SI{8}{\fm}$ with $L \to \infty$ as a reference. We find a consistent
infinite-volume limit, i.e., the $L = \SI{8}{\fm}$ results are almost
indistinguishable from the $L \to \infty$ ones. Visible volume effects start to
appear below $L \lesssim \SI{4}{\fm}$, while for larger system sizes, the
results are very close to one another and basically identical for both boundary
conditions. In general, the crossover lines move towards lower temperatures for
decreasing system sizes. Except for PBC and $L = \SI{3}{\fm}$, the CEPs also
tend to move in direction of higher chemical potentials.

\paragraph{Baryon Number Fluctuations}

Next, we turn to the analysis of baryon number fluctuations around the CEP as
we did in Ref.~\cite{Bernhardt:2022mnx}. Our starting point are the
quark number fluctuations,
\begin{equation}
    \chi_{ijk}^{\upu \upd \ups}
    =
    -
    \frac{1}{T^{4 - (i + j + k)}}
    \,
    \frac{
        \partial^{i + j + k}
        \+
        \Omega
    }{
        \partial \muu^i
        \+
        \partial \mud^j
        \+
        \partial \mus^k
    }
    \,,
\end{equation}
which can be related to the ones for the conserved charges baryon number (B),
electrical charge (Q) and strangeness (S). As a consequence, the corresponding
fluctuations $\chi_{ijk}^{\upB\upQ\upS}$ can be expressed as linear combinations
of quark number fluctuations; see Ref.~\cite{Bernhardt:2022mnx} for details.

Baryon number fluctuations are an interesting quantity to study for a number of
reasons. For instance, they are rather sensitive to and show signatures of phase
transitions and the CEP, and can be related to the moments of the net-baryon
distribution accessible in experiments, e.g., $\chi_3^\upB / \chi_2^\upB =
S_\upB \sigma_\upB$, $\chi_4^\upB / \chi_2^\upB = \kappa_\upB \+ \sigma_\upB^2$,
... Additonally, the fluctuations themselves explicitly depend on the system
volume while their ratios are expected not to. For more information on
fluctuations, see the review~\cite{Luo:2017faz}.

While the QCD grand potential $\Omega$ is inaccessible with DSEs, we can obtain
the quark number fluctuations by means of derivatives of the quark number
densities $n_{f}$ from the quark propagator,
\begin{equation}
    n_{f}
    =
    -
    \frac{
        \partial \Omega
    }{
        \partial \mu_{f}
    }
    \sim
    -
    \frac{T}{L^{3}}
    \sum_{\omega_{m}}
    \sum_{\vec{n} \in \bbZ^{3}}
    \,
    \Tr\bigl[
        \gamma_{4} S_f(\omega_{n} + \ii \mu_{f}, \vec{q}_{\vec{n}})
    \bigr]
    \,.
\end{equation}
In numerical calculations, this expression for the quark number density is
divergent and needs to be regularized. In our finite-volume framework, this
process is somewhat elaborate which is why we refer the reader to
Ref.~\cite{Bernhardt:2022mnx} for details.

\begin{figure*}[t]
    \centering%
    \includegraphics[scale=1.0]{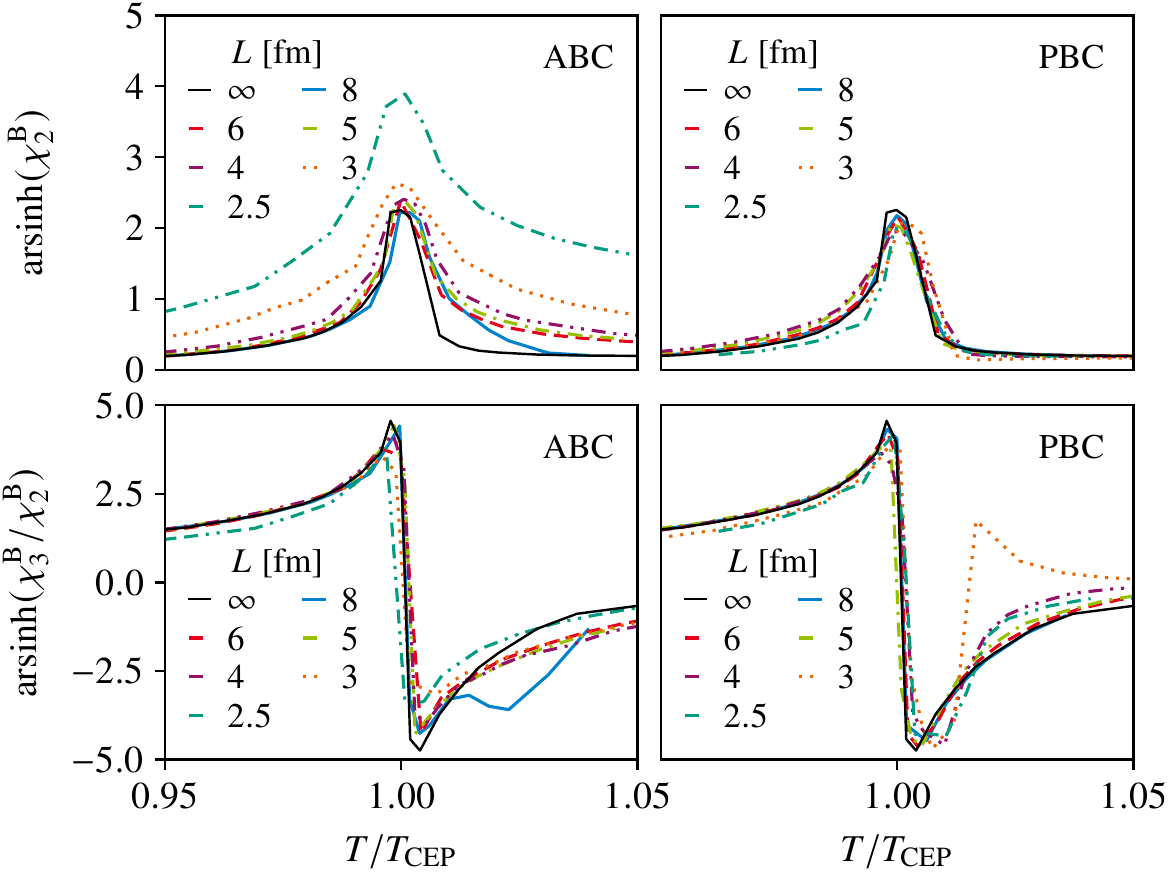}%
    \vspace*{-1mm}%
    \caption{\label{fig:fluctuations}%
        Second order baryon number fluctuations (\emph{top row}) and skewness
        ratio (\emph{bottom row}) against normalized temperature for
        antiperiodic (\emph{left}) and periodic (\emph{right}) boundary
        conditions in different box sizes $L$. All results are obtained at the
        critical chemical potential for the respective system size $\mu =
        \mu_{\textup{CEP}}(L)$.
    }%
\end{figure*}

In Figure~\ref{fig:fluctuations}, we display our results%
\footnote{%
    For better visibility, we display the inverse hyperbolic sine
    ($\mathrm{arsinh}$), which resembles a logarithmic plot in both positive and
    negative direction.%
}
for both the second order baryon number fluctuations (top row) and the skewness
ratio (bottom row) around the CEP for both boundary conditions and system
sizes between $L = \SI{2.5}{\fm}$ and $L = \SI{8}{\fm}$ with $L \to \infty$ for
comparison. The main result we find is that the baryon number fluctuations
themselves show visible volume effects -- especially for ABC -- while the
ratios are indeed basically independent of the system size. This implies that
not only the explicit volume dependence in the ratios cancels but also the
implicit one.

\newpage

\section{Acknowledgements}

We thank Bernd-Jochen Schaefer for fruitful collaboration on the QCD phase diagram in a
finite volume, Ref.~\cite{Bernhardt:2021iql}. 
This work has been supported by the Helmholtz Graduate School for
Hadron and Ion Research for FAIR, the GSI Helmholtzzentrum f\"{u}r
Schwerionenforschung, and the BMBF under Contract No.~05P18RGFCA.

\bibliographystyle{JHEP}
\bibliography{JulianBernhardt_Fairness}

\end{document}